\documentclass[twocolumn]{aastex631}

\usepackage{amsmath}	
\usepackage{amssymb}	
\usepackage{wasysym}
\usepackage{multirow}
\usepackage[flushleft]{threeparttable}
\usepackage{booktabs}

\shorttitle{The HD-TESS catalog}
\shortauthors{Hon et al.}

\graphicspath{{./}{figures/}}

\begin{document}

\title{HD-TESS: An Asteroseismic Catalog of Bright Red Giants within TESS Continuous Viewing Zones}

\correspondingauthor{Marc Hon}
\email{mtyhon@hawaii.edu}

\author[0000-0003-2400-6960]{Marc Hon}
\altaffiliation{NASA Hubble Fellow}
\affiliation{Institute for Astronomy, University of Hawai`i,
2680 Woodlawn Drive, Honolulu, HI 96822, USA}

\author[0000-0002-3322-5279]{James S. Kuszlewicz}  
\affiliation{Center for Astronomy (Landessternwarte), Heidelberg University, K\"onigstuhl 12, 69118 Heidelberg, Germany\\}
\affiliation{Stellar Astrophysics Centre, Department of Physics and Astronomy, Aarhus University, Ny Munkegade 120, DK-8000 Aarhus C, Denmark\\}
\affiliation{Max-Planck-Institut f\"{u}r Sonnensystemforschung, Justus-von-Liebig-Weg 3, 37077 G\"{o}ttingen, Germany}

\author[0000-0001-8832-4488]{Daniel Huber} 
\affiliation{Institute for Astronomy, University of Hawai`i,
2680 Woodlawn Drive, Honolulu, HI 96822, USA}

\author[0000-0002-4879-3519]{Dennis Stello}
\affiliation{School of Physics, The University of New South Wales,
Sydney NSW 2052, Australia}
\affiliation{Stellar Astrophysics Centre, Department of Physics and Astronomy, Aarhus University, Ny Munkegade 120, DK-8000 Aarhus C, Denmark\\}
\affiliation{Sydney Institute for Astronomy (SIfA), School of Physics, University of Sydney,
NSW 2006, Australia}
\affiliation{ARC Centre of Excellence for Astrophysics in Three Dimensions (ASTRO-3D), Australia}

\author{Claudia Reyes}
\affiliation{School of Physics, The University of New South Wales,
Sydney NSW 2052, Australia}

\begin{abstract}
We present HD-TESS, a catalog of 1,709 bright ($V\sim3-10$) red giants from the Henry Draper (HD) Catalog with asteroseismic measurements based on photometry from NASA's Transiting Exoplanet Survey Satellite (TESS). Using light curves spanning at least six months across a single TESS observing cycle, we provide measurements of global asteroseismic parameters ($\nu_{\mathrm{max}}$ and $\Delta\nu$) and evolutionary state for each star in the catalog. We adopt literature values of atmospheric stellar parameters to estimate the masses and radii of the giants in our catalog using asteroseismic scaling relations, and observe that HD-TESS giants on average have larger masses compared to \textit{Kepler} red giants. Additionally, we present the discovery of oscillations in 99 red giants in astrometric binary systems, including those with subdwarf or white dwarf companions. Finally, we benchmark radii from asteroseismic scaling relations against those measured using long-baseline interferometry for 18 red giants and find that correction factors to the scaling relations improve the agreement between asteroseismic and interferometric radii to approximately 3\%.



\end{abstract}





\section{Introduction} \label{sec:intro}
Bright stars are amongst some of the most well-studied celestial objects. They are prime targets for a myriad of studies, including high-resolution spectroscopic surveys of the solar neighbourhood (e.g., \citealt{Luck_2007, Hekker_2007, Soubiran_2010, Feuillet_2018}) as well as planet searches using the Doppler velocity method (e.g., \citealt{Pepe_2011, Addison_2019}). Furthermore, fundamental measurements such as angular diameters (e.g., \citealt{Baines_2010, vonBraun_2017}), and dynamical measurements of visual binaries  (e.g., \citealt{McAlister_1977, Retterer_1982, Mason_2001}) are limited to only bright stars. Some of these stars additionally serve as benchmarks for studying stellar populations in the Galaxy \citep{Jofre_2018}, and therefore a precise characterization of their properties is essential. 

Asteroseismology, the study of stellar oscillations, can provide precise measurements of the masses and radii of Sun-like stars \citep{JCD_1984, Bedding_2014, Garcia_2015, Chaplin_2014, Hekker_2017}. Over the past decade, the availability of precise, continuous photometry from space missions CoRoT \citep{Baglin_2009}, \textit{Kepler} \citep{Borucki_2010} and \textit{K2} \citep{Howell_2014} has enabled the seismic characterization of the fundamental properties of thousands of red giants across distant stellar populations within the Milky Way \citep{Miglio_2009, Pinsonneault_2018, Yu_2018}, which has been pivotal in studies of Galactic archeology (e.g., \citealt{Silva-Aguirre_2018, Zinn_2020, Miglio_2021}). However, the asteroseismology of bright ($V \apprle7$) stars has previously been limited to only several dozens of targets that can be measured from ground-based radial velocity surveys (e.g., \citealt{Kjeldsen_1995, Frandsen_2002, Malla_2020}), or a small subset of \textit{Kepler/K2} stars for which smear and halo photometry can be applied \citep{Pope_2016, White_2017, Pope_2019}.
It is only through recent space-borne photometric surveys specifically targeting bright stars such as the Bright Target Explorer \citep[BRITE]{Weiss_2014} and NASA's Transiting Exoplanet Survey Satellite \citep[TESS]{Ricker_2015} that the asteroseismology of a large ensemble of bright stars is now possible.

In this paper, we aim to characterize the fundamental parameters of bright red giants observed by TESS. The prime targets for this analysis are stars near or within the TESS Continuous Viewing Zones (CVZs), which are parts of the sky with maximal overlap between the $24^{\circ}\times96^{\circ}$ TESS observing fields (sectors) within Cycles 1-3 of the mission. The long photometric time series resulting from this overlap lead to a greater frequency resolution in the Fourier domain, which in turn improves the precision of asteroseismic measurements. As a result, we aim to produce a catalog of seismically-determined masses and radii for bright red giants that will be useful for informing other studies involving giant stars.
This catalog also seeks to complement previous asteroseismic studies of red giants with TESS \textemdash{} specifically the work by \citet{Silva-Aguirre_2020} that measured the oscillation properties of 25 bright red giants, the Galactic archeology study by \citet{Mackereth_2021} that seismically analyzed $\sim$5,500 red giants in the southern CVZ, and the near all-sky detection and seismic characterization of $\sim$158,000 giants by \citet{Hon_2021}. In contrast to these studies, we will focus on the bright stars in both CVZs by basing our selection on red giants from the Henry Draper (HD) catalog \citep{Cannon_1993, Nesterov_1995}. The cross-match between our selection from the HD catalog with TESS observations thus yields the HD-TESS catalog whose construction we detail in the following.

\section{Data}

\begin{figure}
    \centering
	\includegraphics[width=1.\columnwidth]{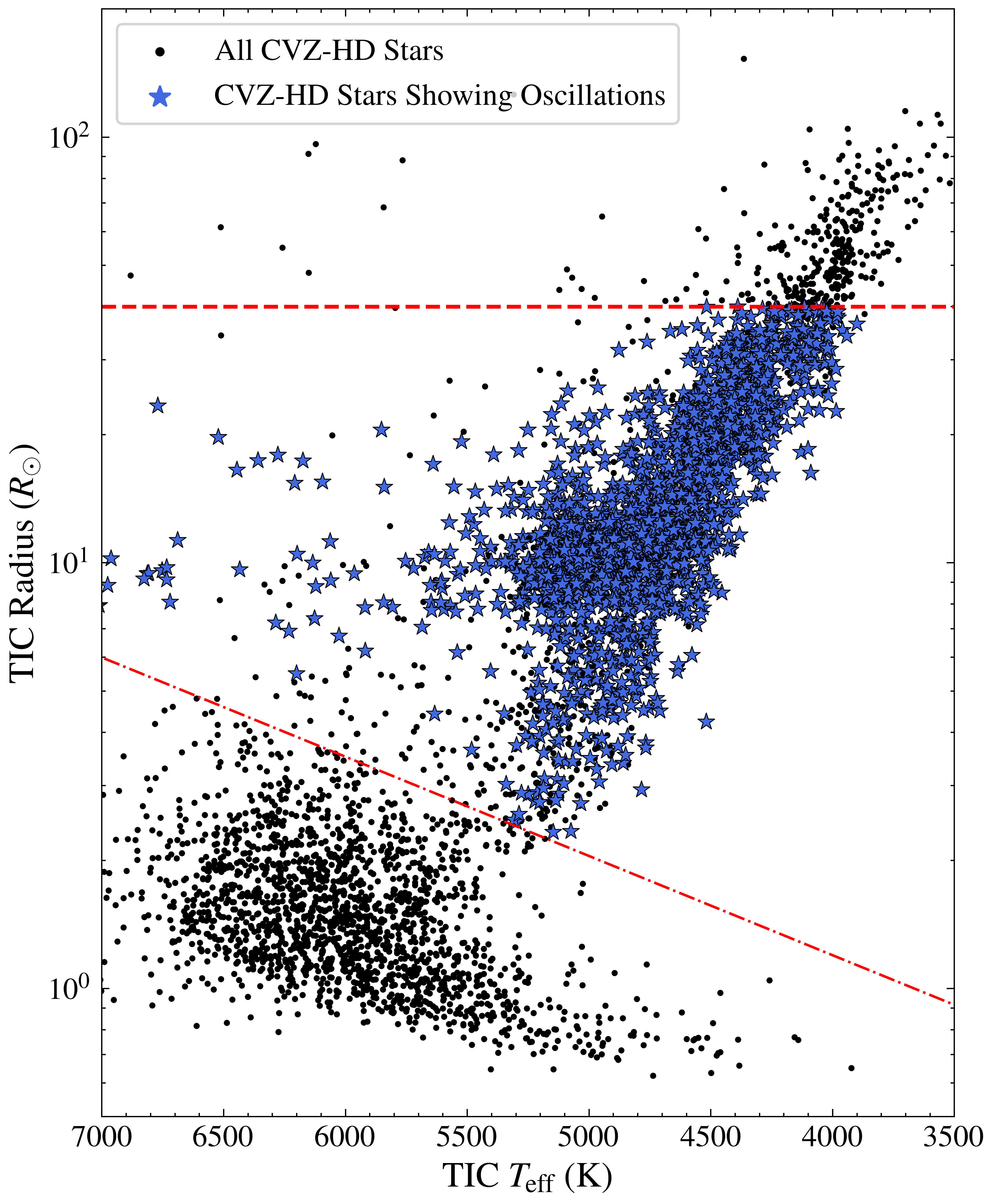}
    \caption{Stellar radius versus effective temperature for stars with spectral type G-K from the Henry Draper catalog that have been observed for at least six sectors in Cycles 1+2 from TESS and therefore lie near or within the TESS Continuous Viewing Zones (CVZs). The temperatures and radii in this plot are from the TESS Input Catalog \citep[TIC]{Stassun_2019}. Stars in our sample are those that lie between the two lines in this diagram whereby the upper line corresponds to a $40R_{\odot}$ radius cut-off, while the lower line corresponds to a fiducial boundary separating early subgiants from the base of the red giant branch (see text). The blue points correspond to stars that show a power excess corresponding to red giant oscillations in their power spectrum.}
    \label{fig:hr_selection}
\end{figure}

There are two sources of light curves in this study. The first is the TESS Science Processing Operations Center pipeline \citep[SPOC]{Twicken_2016, Jenkins_2020} obtained from the Mikulski Archive for Space Telescopes (MAST) at the Space Telescope Science Institute, accessed via\dataset[doi:10.17909/t9-nmc8-f686]{https://doi.org/10.17909/t9-nmc8-f686}. The second is the MIT Quick-Look Pipeline \citep[QLP]{Huang_2020a, Huang_2020b}, accessed via\dataset[doi:10.17909/t9-r086-e880]{https://doi.org/10.17909/t9-r086-e880}. Generally, we find that SPOC light curves better preserve flux variability corresponding to giant oscillations for bright and saturated targets with T$_{\mathrm{mag}}\apprle7$ compared to those from the MIT Quick-Look Pipeline \citep[QLP]{Huang_2020a, Huang_2020b}. Many of the fainter HD stars in our sample, however, were not targeted by the SPOC pipeline but have light curves generated from the QLP. Because the data from QLP's simple aperture photometry is generally adequate for observing giant oscillations at fainter magnitudes (\citealt{Hon_2021}), our approach here is to prioritize SPOC light curves whenever possible, otherwise we default to QLP photometry.

Our analysis includes observations up to Sector 39 for stars with SPOC light curves, such that stars near the southern ecliptic poles will have up to two years of data as of the end of the year 2021.
For stars having only QLP photometry, we use data across Cycles 1 and 2, meaning the maximum length of the time series for such stars in this work will be one year. Although QLP light curves across Cycle 3 are available \citep{Kunimoto_2019}, the data have a reduced sampling cadence of 10 minutes as opposed to the 30-minute cadence across the first two TESS Cycles. To simplify the treatment of data in this study, we incorporate a consistent observing cadence for each star's light curve and therefore delegate the task of matching difference observing cadences as part of future work.

\begin{figure*}
    \centering
	\includegraphics[width=2.\columnwidth]{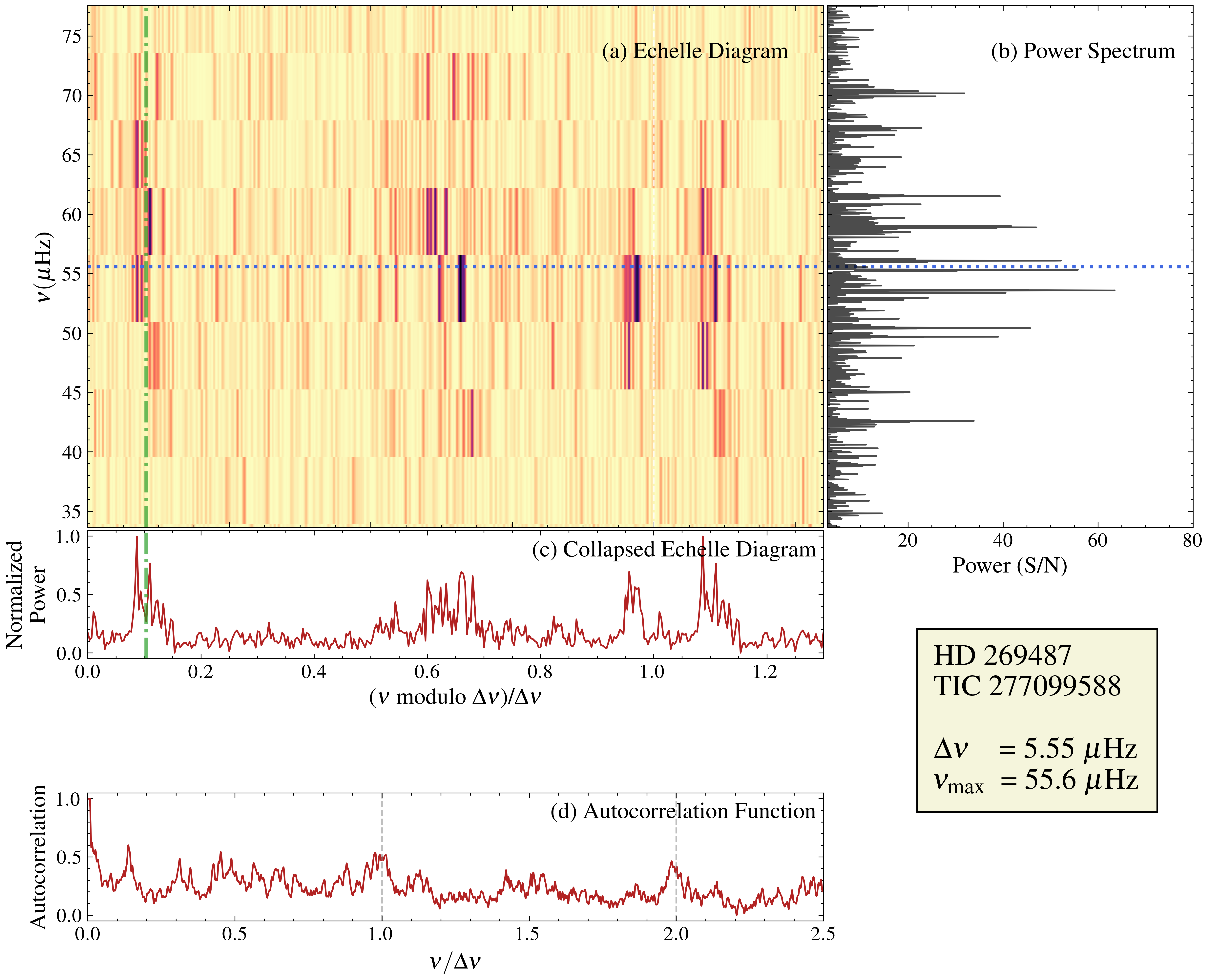}
    \caption{Diagnostic plot for HD 269847 (TIC 277099588) showing (a) an \'echelle diagram of the (b) oscillation spectrum of the star, (c) the collapsed \'echelle diagram computed as the summation of power across each row in the \'echelle diagram, and (d) the autocorrelation function, constructed by convolving the oscillation spectrum in (b) with itself. The \'echelle diagrams are replicated up to ($\nu~$modulo $\Delta\nu$/)$\Delta\nu$ = 1.25 to better visualize the repetition of the ridges across radial orders.
    The blue horizontal line in panels (a) and (b) corresponds to $\nu_{\mathrm{max}}$, while the green vertical line in panels (a) and (c) corresponds to the position of the radial mode-ridge in the \'echelle diagrams, also known as the offset $\epsilon$. The vertical lines in panel (d) correspond to spacings of 1 and 2 $\Delta\nu$, and is where peaks are expected if there is strong regularity in the oscillation spectrum given a candidate $\Delta\nu$ value.}
    \label{fig:vetting}
\end{figure*}

\section{Target Selection}\label{target_selection}
First, we identify 5,711 stars with spectral types G and K from the HD catalog that have been observed for at least six sectors in TESS Cycles 1 and 2. Next, we impose a selection based on a star's effective temperature and radius from the TESS Input Catalog version 8.2 \citep[TIC]{Stassun_2019}, as follows:
\begin{itemize}
    \item $R < 40~R_{\odot}$ as an upper boundary to filter out long-period variables.
    \item $R \geq 10^p~R_{\odot}$, with $p= ( \frac{T_{\mathrm{eff}}\mathrm{(K)}}{7000}- \frac{3}{7}) \log\big(\frac{300}{7}\big) + \log(0.7)$. This is the boundary used by \citet{Hon_2019} to discriminate early subgiant stars from those beginning to ascend the red giant branch.
\end{itemize}

This selection is depicted in Figure \ref{fig:hr_selection}. Under this selection, we reduce the initial sample of stars to 3,142, from which 2,611 show red giant oscillations. Note that 2,042 of the 2,611 stars are already known to show oscillations based on previous work by \citet{Hon_2021}
This implies that 569 red giants in this sample are new inclusions that were not previously detected from one month of TESS data.

\subsection{Asteroseismic Measurements}\label{section:asteroseismic_measurements}
We extract initial estimates for the frequency at maximum oscillation power ($\nu_{\mathrm{max}}$), and the large frequency separation ($\Delta\nu$) for the sample of 2,611 giants using the SYD asteroseismic data pipeline \citep{Huber_2009}. With improvements to the SYD pipeline as specified in \citet{Huber_2011} and \citet{Yu_2018}, the pipeline has previously been applied for the asteroseismology of shorter time series including those from \textit{K2} \citep{Zinn_2020, Zinn_2022} and even from TESS itself \citep{Stello_2021}. Beginning with the SYD pipeline's $\nu_{\mathrm{max}}$ estimates, we further refine the estimates by performing a Bayesian fit using a model comprising a Gaussian envelope with three background components to the star's power spectrum as described by \citet{Themessl_2020}. The resulting $\nu_{\mathrm{max}}$ from this final fit is our reported estimate.

The raw output from the SYD pipeline reports a $\Delta\nu$ measured automatically from the autocorrelation function of the oscillation power spectrum for each input star. However, complicating features in the frequency domain such as lower signal-to-noise ratios, additional mode structure from dipole mixed modes (e.g., \citealt{Bedding_2011}), or the overlap of other periodic signals unrelated to oscillations (e.g., \citealt{Colman_2017}) may result in inaccurate measurements of $\Delta\nu$. To avoid such occurrences, we first vet the SYD pipeline outputs using the neural network classifier introduced by \citet{Reyes_2022}. Briefly, the classifier takes as input extracted features across a variety of commonly used visual diagrams in asteroseismology \textemdash{} such as the autocorrelation plot and the echelle diagram of the oscillation modes \textemdash{} and subsequently outputs a probability indicating whether the $\Delta\nu$ measurement used to construct such diagrams corresponds to a valid measurement. To further ensure the robustness of the $\Delta\nu$ measurements, we additionally inspect the oscillation spectrum of each star using diagnostic plots such as shown in Figure \ref{fig:vetting}. We perform this inspection in addition to the automatic vetting by the neural network because it allows us to recover a fraction of initially rejected stars that have $\Delta\nu$ erroneously estimated by the seismic pipeline (and thus rightfully rejected by the automatic vetting). To recover $\Delta\nu$ for such stars, we manually tune $\Delta\nu$ such that: (1) vertical ridges are formed in the \'echelle diagram; (2) peaks are observed at 1 and 2 $\Delta\nu$ in the autocorrelation of the oscillation spectrum, as indicated in Figure \ref{fig:vetting}. The uncertainty in $\Delta\nu$ is taken as the shift required to disrupt the vertical alignment of the ridges in the \'echelle diagram and is approximately at the 2-3\% level.
We additionally inspect other stars that pass the automatic vetting to ensure that the same visual criteria are met.

\begin{figure}
    \centering
	\includegraphics[width=1.\columnwidth]{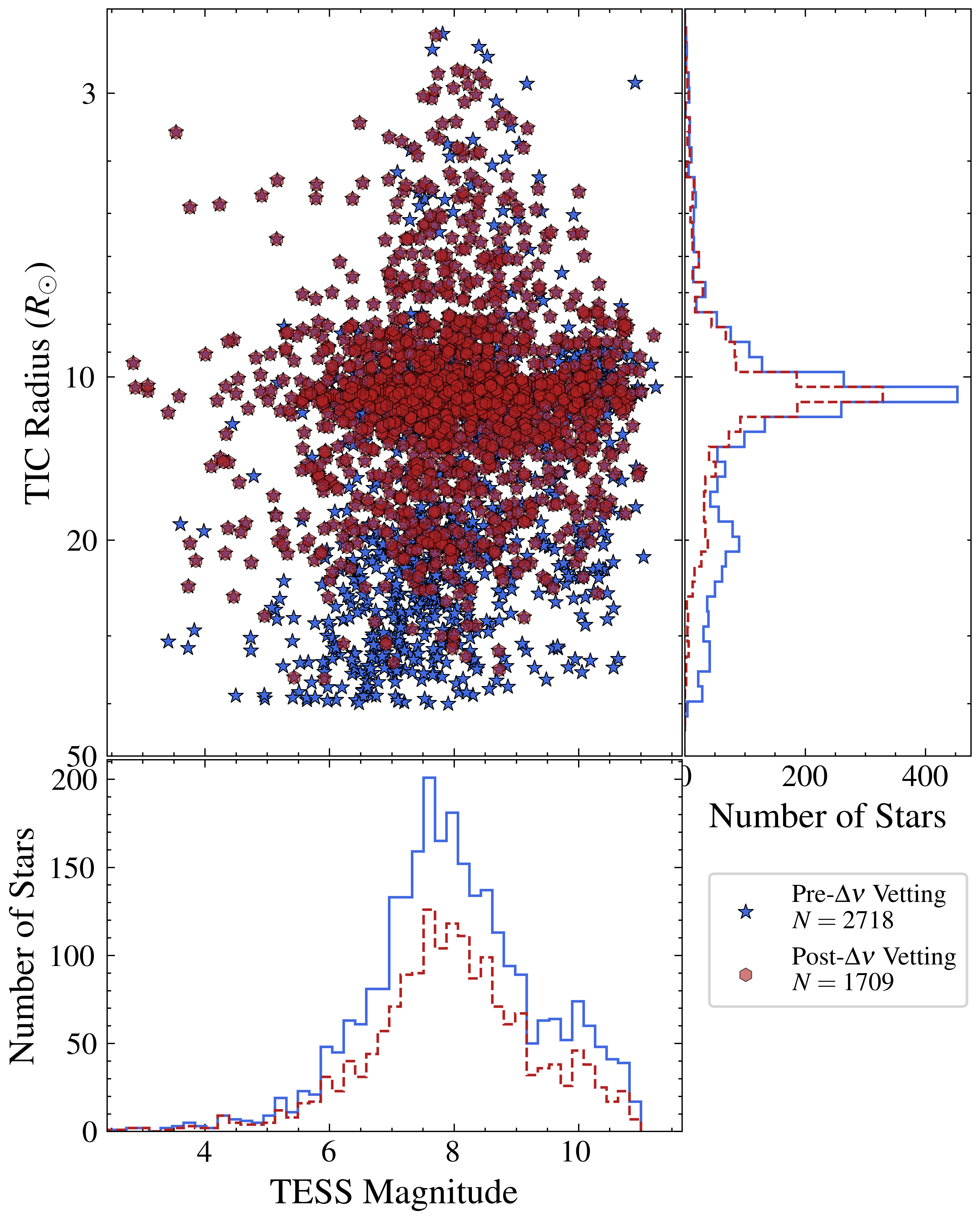}
    \caption{(Left) Radius-magnitude distribution of oscillating giants before (blue) and after (red) vetting of their large frequency separations $\Delta\nu$ as described in Section \ref{section:asteroseismic_measurements}. Similar to Figure \ref{fig:hr_selection}, the radii used in the plot are adopted from the TESS Input Catalog. The plot is aligned such that more luminous giants (with smaller $\nu_{\mathrm{max}}$) are lower, consistent with how the $\nu_{\mathrm{max}}$-magnitude distribution for giants are commonly visualized (e.g., \citealt{Stello_2017, Zinn_2020, Hon_2021}).}
    \label{fig:pre_post}
\end{figure}

\begin{figure*}
    \centering
	\includegraphics[width=2.\columnwidth]{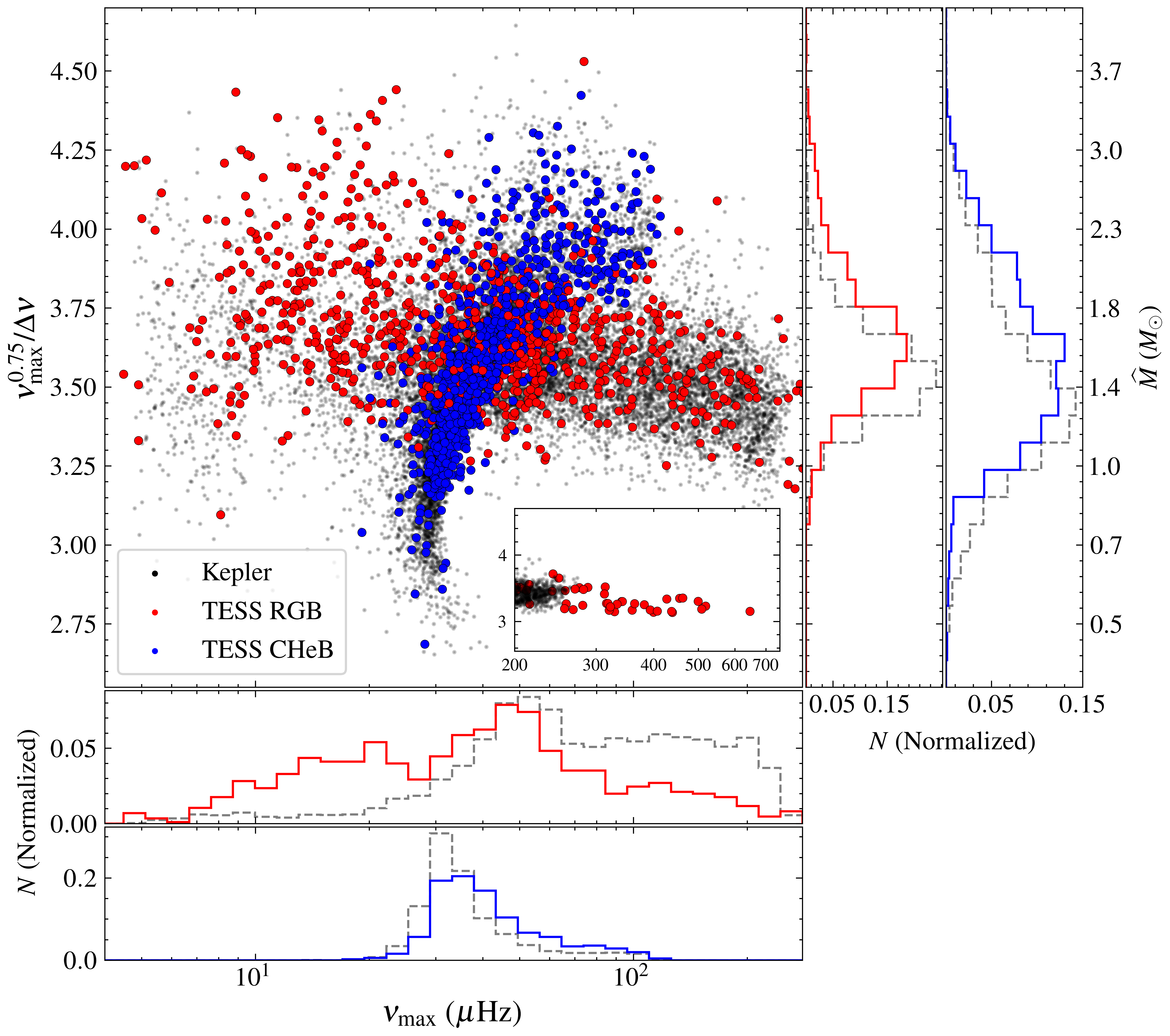}
    \caption{Asteroseismic measurements for the 1,709 stars in the HD-TESS catalog. On the abscissa is frequency at maximum oscillation power ($\nu_{\mathrm{max}}$), while the ordinate $\nu_{\mathrm{max}}^{0.75}/\Delta\nu\sim M^{0.25}T_{\mathrm{eff}}^{-0.375}$ is an asteroseismic proxy for stellar mass.
    The quantity $\widehat{M}$ is the mass proxy computed assuming $T_{\mathrm{eff}}=5000$K, shown here to provide an approximation of range of red giant masses reported in this work.
    In red are hydrogen shell-burning giants ascending the giant branch (RGB), while in blue are core helium-burning giants (CHeB). 
    For comparison, corresponding measurements for \textit{Kepler} red giants \citep{Yu_2018} are shown in gray. The inset shows measurements for low-luminosity TESS giants in our sample.}
    \label{fig:nike}
\end{figure*}

The result of the vetting is shown in Figure \ref{fig:pre_post}. 
From 2,611 oscillating giants, we retain 1,709 for which their $\Delta\nu$ can be identified with a high level of confidence. Most giants filtered out in this vetting stage are those with $R\apprge20R_{\odot}$ (corresponding to $\nu_{\mathrm{max}}\apprle10\mu$Hz) and therefore have long-period oscillations for which frequency separations are typically difficult to measure with one to two years of data. We also observe a sharp cutoff corresponding to an absence of faint stars with small radii, where stars beyond this limit show too low signal-to-noise to have detectable oscillations \citep{Stello_2017, Hon_2021, Stello_2022}.

Finally, we categorize the giants in our sample as either those that are hydrogen shell-burning while ascending the red giant branch (RGB) or core helium-burning stars within the red/secondary clump (CHeB) by applying the \citet{Hon_2017, Hon_2018} deep learning classifier, which uses the frequency distribution of oscillation modes within a star's collapsed \'echelle diagram (e.g., Figure \ref{fig:vetting}c) to determine an evolutionary state. The 1,709 stars with $\Delta\nu$, $\nu_{\mathrm{max}}$, and evolutionary state determinations thus forms the HD-TESS catalog and are tabulated in Table \ref{table:seismic_measurements}. The $\nu_{\mathrm{max}}$ measurements of all 2,611 oscillating giants identified in this study, including those whose $\Delta\nu$ cannot be reliably measured, is tabulated in Appendix \ref{section:appendix_numax}. 

\section{Fundamental Stellar Parameters}
\subsection{Seismic Measurements}

\begin{deluxetable*}{cccccccc}[t]
\tablenum{1}
\tablecaption{Asteroseismic quantities for 1,709 stars in the HD-TESS catalog. Quantities $\kappa_M$ and $\kappa_R$ correspond respectively to mass and radius `coefficients', which are convenient quantities for estimating masses and radii using the asteroseismic scaling relations (Equations \ref{mass_scaling_relation} and \ref{radius_scaling_relation}). The evolutionary phase of each giant is provided as either hydrogen shell-burning (RGB) or core helium-burning (CHeB). The full version of this table is available in a machine-readable format in the online journal, with a portion shown here for guidance regarding its form and content. \label{table:seismic_measurements}}
\tablewidth{0pt} 
\tablehead{ 
\colhead{HD number} & \colhead{TIC} & \colhead{T$_\mathrm{mag}$} & \colhead{$\Delta\nu$} & \colhead{$\nu_{\mathrm{max}}$} 
& \colhead{$\kappa_M$} & \colhead{$\kappa_R$} & \colhead{Phase}
\\
\colhead{} & \colhead{} & \colhead{(mag)} & \colhead{($\mu$Hz)} & \colhead{($\mu$Hz)} &  \colhead{} & \colhead{}  & \colhead{}
}
\startdata
22349 & 31852879 & 8.6 & $3.82\pm0.04$ & $34.04\pm0.50$& $2.10\pm0.13$& $13.79\pm0.35$ & RGB\\
23128 & 238180822 & 6.6 & $3.05\pm0.04$ & $25.24\pm0.39$& $2.10\pm0.15$& $16.03\pm0.49$ & RGB\\
23381 & 31960832 & 7.5 & $3.59\pm0.02$ & $30.97\pm0.50$& $2.02\pm0.11$& $14.19\pm0.28$ & CHeB\\
23399 & 425997943 & 7.9 & $4.18\pm0.04$ & $34.24\pm1.39$& $1.48\pm0.19$& $11.58\pm0.52$ & CHeB\\
$\cdots$&$\cdots$&$\cdots$&$\cdots$&$\cdots$&$\cdots$&$\cdots$&$\cdots$\\
311740 & 271795748 & 8.4 & $7.38\pm0.02$ & $80.52\pm0.78$& $1.99\pm0.06$& $8.73\pm0.10$ & RGB\\
370635 & 370115759 & 10.9 & $2.85\pm0.02$ & $22.95\pm0.35$& $2.07\pm0.11$& $16.7\pm0.35$ & RGB\\
\enddata
\end{deluxetable*}

Figure \ref{fig:nike} shows the results of our seismic measurements, where we plot $\nu_{\mathrm{max}}$ against the quantity $\nu_{\mathrm{max}}^{0.75}/\Delta\nu\sim M^{0.25}T_{\mathrm{eff}}^{-0.375}$, which acts as a temperature ($T_{\mathrm{eff}}$)-dependent  proxy for stellar mass ($M$). With TESS observations spanning at least 6 months within the CVZs, we are able to clearly distinguish between RGB and CHeB populations as predicted by \citet{Mosser_2019}. Other population-level features that are well-resolved from the figure includes the sharpness of the right edge of the CHeB distribution that marks the zero-age CHeB phase (e.g., \citealt{Kallinger_2010, Li_2021_edge}) as well as an overdensity of RGB stars at $\sim50\mu$Hz indicating the luminosity bump in first ascent giants \citep{Refsdal_1970,JCD_2015, Khan_2018}. Additionally, we observe a systematic increase in the stellar mass proxy for both RGB and CHeB TESS populations compared to the giants from \textit{Kepler} \citep[black points]{Yu_2018}, which we discuss further in Section \ref{section:mass_radii}.

Estimates of the mass ($M$) and radius ($R$) of each giant can be obtained using the following asteroseismic scaling relations:

\begin{align}
    \nonumber\frac{M}{M_{\odot}} & = \bigg(\frac{\Delta\nu}{f_{\Delta\nu}\Delta\nu_{\odot}} \bigg)^{-4}\bigg(\frac{\nu_{\mathrm{max}}}{\nu_{\mathrm{max, \odot}}}\bigg)^3\bigg(\frac{T_{\mathrm{eff}}}{T_{\mathrm{eff, \odot}}}\bigg)^{3/2}\\\label{mass_scaling_relation}
    & = \frac{\kappa_M}{f_{\Delta\nu}^{-4}}~\bigg(\frac{T_{\mathrm{eff}}}{T_{\mathrm{eff, \odot}}}\bigg)^{3/2},\\
    \nonumber\frac{R}{R_{\odot}} &= \bigg(\frac{\Delta\nu}{f_{\Delta\nu}\Delta\nu_{\odot}} \bigg)^{-2}\bigg(\frac{\nu_{\mathrm{max}}}{\nu_{\mathrm{max, \odot}}}\bigg)\bigg(\frac{T_{\mathrm{eff}}}{T_{\mathrm{eff, \odot}}}\bigg)^{1/2}.\\\label{radius_scaling_relation}
     &= \frac{\kappa_R}{f_{\Delta\nu}^{-2}}~\bigg(\frac{T_{\mathrm{eff}}}{T_{\mathrm{eff, \odot}}}\bigg)^{1/2}.    
\end{align}
\\
These relations are a consequence of the approximate proportionality of $\Delta\nu$ to the square root of stellar mean density \citep{Ulrich_1986} as well as that of $\nu_{\mathrm{max}}$ to the star's acoustic cutoff frequency \citep{Brown_1991, Kjeldsen_1995, Belkacem_2011}. The quantity $f_{\Delta\nu}$ is a correction term used to correct for any deviations from the proportionality assumption of the former relation, and has been shown to provide better agreement between independently measured fundamental parameters from eclipsing binaries \citep{Gaulme_2016, Brogaard_2018, Kallinger_2018} and \textit{Gaia} parallaxes \citep{Huber_2017, Hall_2019, Zinn_2019}. Here, we adopt the solar reference values of $\Delta\nu_{\odot}=135.1\mu$Hz, $\nu_{\mathrm{max, \odot}}=3090\mu$Hz, and $T_{\mathrm{eff, \odot}}=5777~$K \citep{Huber_2011, Huber_2013}.

Following \citet{Sharma_2016}, we condense the scaling relations into a more compact form by defining the coefficients $\kappa_M = (\Delta\nu/\Delta\nu_{\odot})^{-4}\cdot (\nu_{\mathrm{max}}/\nu_{\mathrm{max, \odot}})^{3}$ for mass and $\kappa_R = (\Delta\nu/\Delta\nu_{\odot})^{-2}\cdot (\nu_{\mathrm{max}}/\nu_{\mathrm{max, \odot}})$ for radius to represent the seismic aspect of both relations. Similar to \citet{Zinn_2020}, the purpose of reporting such a representation is to provide the user with the choice of using their own atmospheric parameters to estimate masses and radii. Not only do the scaling relations have an explicit temperature dependence, but the correction term $f_{\Delta\nu}$ is also often computed with reference to stellar models as a function of mass, temperature, metallicity, and evolutionary phase of the star \citep{Sharma_2016, Serenelli_2017, Rodrigues_2017, Li_2022}. To obtain cursory estimates of masses and radii in Section \ref{section:mass_radii}, we adopt the \citet{Sharma_2016} correction.

\begin{figure}
    \centering
	\includegraphics[width=1.\columnwidth]{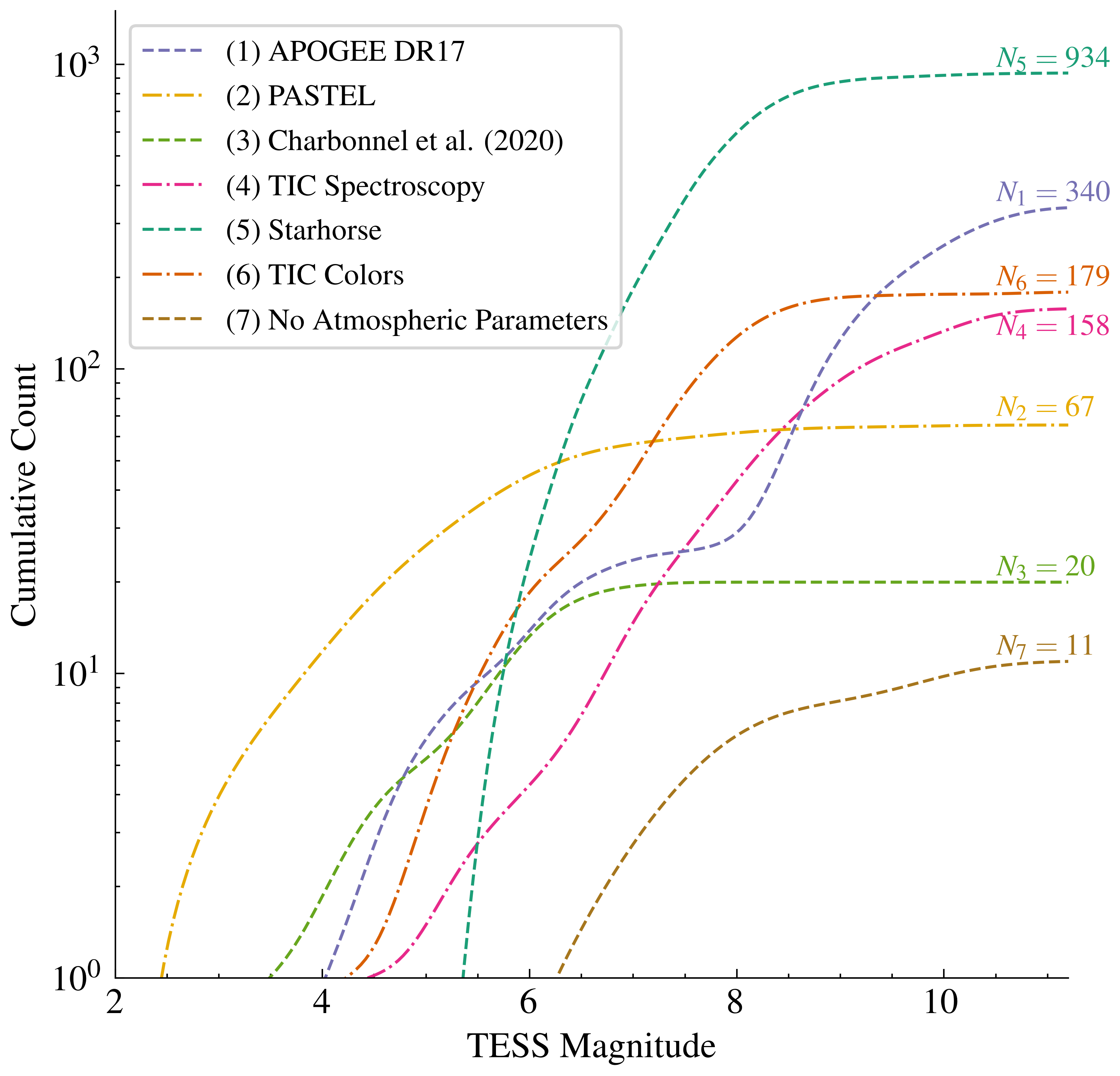}
    \caption{Cumulative histograms of atmospheric parameters sources for HD-TESS giants as a function of TESS magnitude. The number associated with each source in the legend indicates its priority in the selection, while the quantity $N$ enumerates the total number of stars from that particular source. Sources 1-4 provide measurements using spectroscopy, while sources 5 and 6 apply spectrophotometric methods and color-photometry relations, respectively.}
    \label{fig:spect_sources}
\end{figure}

\subsection{Estimating Masses and Radii} \label{section:mass_radii}
A uniform estimation of masses and radii using Equations \ref{mass_scaling_relation} and \ref{radius_scaling_relation} for all stars in our sample is not a straightforward task due to the lack of a homogeneous source of atmospheric parameters for such stars. We therefore collect atmospheric parameters from a variety of catalogs as follows, in decreasing priority order:

\begin{enumerate}
    \item The SDSS/APOGEE-2 Data Release 17 catalog \citep{Abdurrouf_2022}.
    \item The PASTEL catalog \citep{Soubiran_2020}.
    \item Spectroscopic parameters from \citet{Charbonnel_2020} 
    \item Spectroscopic parameters from the TESS Input Catalog, which adopts such measurements across a variety of other spectroscopic catalogs (see \citealt{Stassun_2018, Stassun_2019}).
    \item Spectrophotometric temperatures and metallicities from the \textit{Starhorse} Bayesian inference tool \citep{Queiroz_2018} based on \textit{Gaia} EDR3 data \citep{Anders_2022}.
    \item Photometric temperatures from the TESS Input Catalog, which are determined using the color-$T_{\mathrm{eff}}$ relations outlined in \citet[their Section 2.3.4]{Stassun_2019}. Stars in this category have no metallicity information.
\end{enumerate}

A breakdown of the adopted atmospheric parameters from each catalog is shown in Figure \ref{fig:spect_sources}. Notably, spectroscopic surveys (catalogs 1-4) focus primarily on bright targets with T$_{\mathrm{mag}} \apprle$ 6 or on fainter targets with T$_{\mathrm{mag}} \apprge$ 9, and therefore lack coverage for intermediate brightness stars within our sample. 
It is for this reason that we also include atmospheric parameters from spectrophotometric or photometric catalogs for completeness, despite their lack of precision as opposed to spectroscopic measurements (e.g., Figure 18 from \citealt{Anders_2022}). We adopt the reported temperature and metallicity uncertainties from each catalog whenever available. For stars with multiple measurements in the PASTEL catalog, we adopt the standard deviation between the measurements as the uncertainty. Across all atmospheric parameter catalogs, we assign a minimum uncertainty of 0.05 dex for metallicity and 2\% for temperature \citep{Tayar_2022}.

\begin{deluxetable*}{ccccccc}[t]
\tablenum{2}
\tablecaption{Masses and radii for 1,709 red giants in the HD-TESS catalog. The source of atmospheric parameters adopted for each star is indicated by the following: (1)\textemdash{} SDSS/APOGEE-2 DR17 \citep{Abdurrouf_2022}, (2)\textemdash{} PASTEL \citep{Soubiran_2020}, (3)\textemdash{} \citealt{Charbonnel_2020}, (4)\textemdash{} TESS Input Catalog \citep[TIC]{Stassun_2019} spectroscopy, (5)\textemdash{} \textit{Starhorse} \textit{Gaia} EDR3 \citep{Queiroz_2018, Anders_2022}, (6)\textemdash{} TIC photometric colors, (7)\textemdash{} No atmospheric parameters. The full version of this table is available in a machine-readable format in the online journal, with a portion shown here for guidance regarding its form and content. \label{table:mass_radii_measurements}}
\tablewidth{0pt} 
\tablehead{ 
\colhead{HD number} & \colhead{TIC} & \colhead{$T_{\mathrm{eff}}$} & \colhead{[M/H]} & \colhead{$M$} 
& \colhead{$R$} & \colhead{Atmospheric Parameter Source}
\\
\colhead{} & \colhead{} & \colhead{(K)} & \colhead{(dex)} & \colhead{($M_{\odot}$)} &  \colhead{($R_{\odot}$)} & \colhead{}
}
\startdata
22349 & 31852879 &\hspace{5pt}$4859\pm132$ & $-0.20\pm0.30$ & $1.50\pm0.14$& $12.18\pm0.49$& 5\\
23128 & 238180822 & $4606\pm92$ & $-0.10\pm0.25$ & $1.34\pm0.11$& $13.50\pm0.49$& 5\\
23381 & 31960832 & $4612\pm92$ &\hspace{7pt}$0.20\pm0.15$ & $1.45\pm0.10$& $12.71\pm0.31$& 5\\
23399 & 425997943 &\hspace{5pt}$4816\pm131$ &\hspace{7pt}$0.30\pm0.30$ & $1.13\pm0.16$& $10.54\pm0.54$& 5\\
$\cdots$&$\cdots$&\hspace{10pt}$\cdots$&\hspace{8pt}$\cdots$&$\cdots$&\hspace{7pt}$\cdots$&$\cdots$\\
311740 & 271795748 & $4830\pm97$ & $-0.31\pm0.24$ & $1.38\pm0.09$& \hspace{5pt}$7.61\pm0.20$& 5\\
370635 & 370115759 &$4403\pm88$ & $-0.06\pm0.05$& $1.21\pm0.08$& $13.63\pm0.37$ & 1\\
\enddata
\end{deluxetable*}

\begin{figure}
    \centering
	\includegraphics[width=1.\columnwidth]{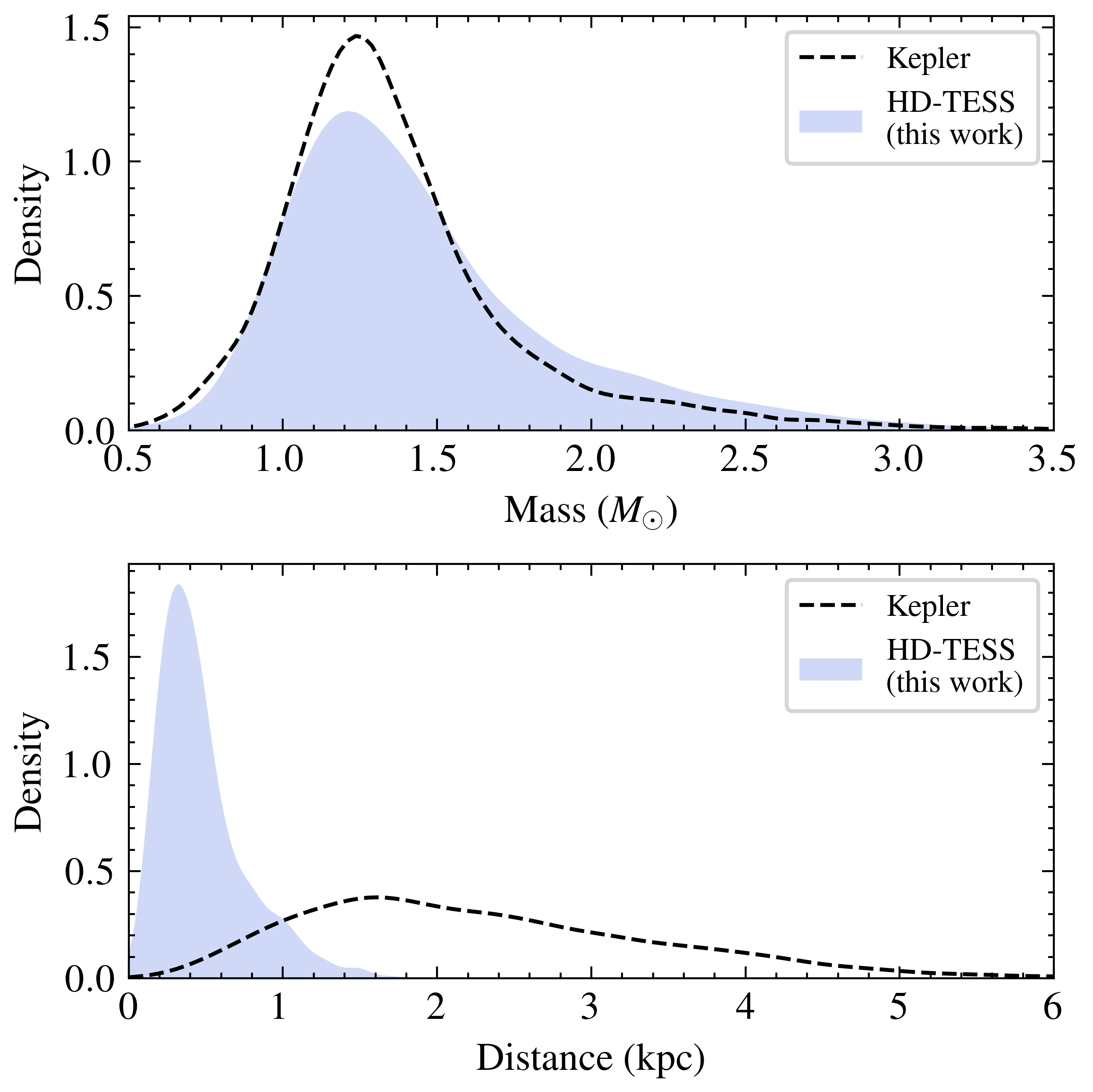}
    \caption{(Top) Kernel density estimates comparing the masses of HD-TESS giants (shaded) with those from a population of giants observed by \textit{Kepler} \citep{Yu_2018}. Compared to the \textit{Kepler} giants, our TESS sample has a greater fraction of stars above 1.5$M_{\odot}$, which is consistent with comparisons of the seismic mass proxy $\nu_{\mathrm{max}}^{0.75}/\Delta\nu$ between both samples in Figure \ref{fig:nike}. (Bottom) A comparison of distances between the two giant star samples. TESS probes stellar populations mainly within the solar neighbourhood, whereas \textit{Kepler} probes populations much farther into the Milky Way disk.}
    \label{fig:kde_mass}
\end{figure}

\begin{figure*}
    \centering
	\includegraphics[width=2.\columnwidth]{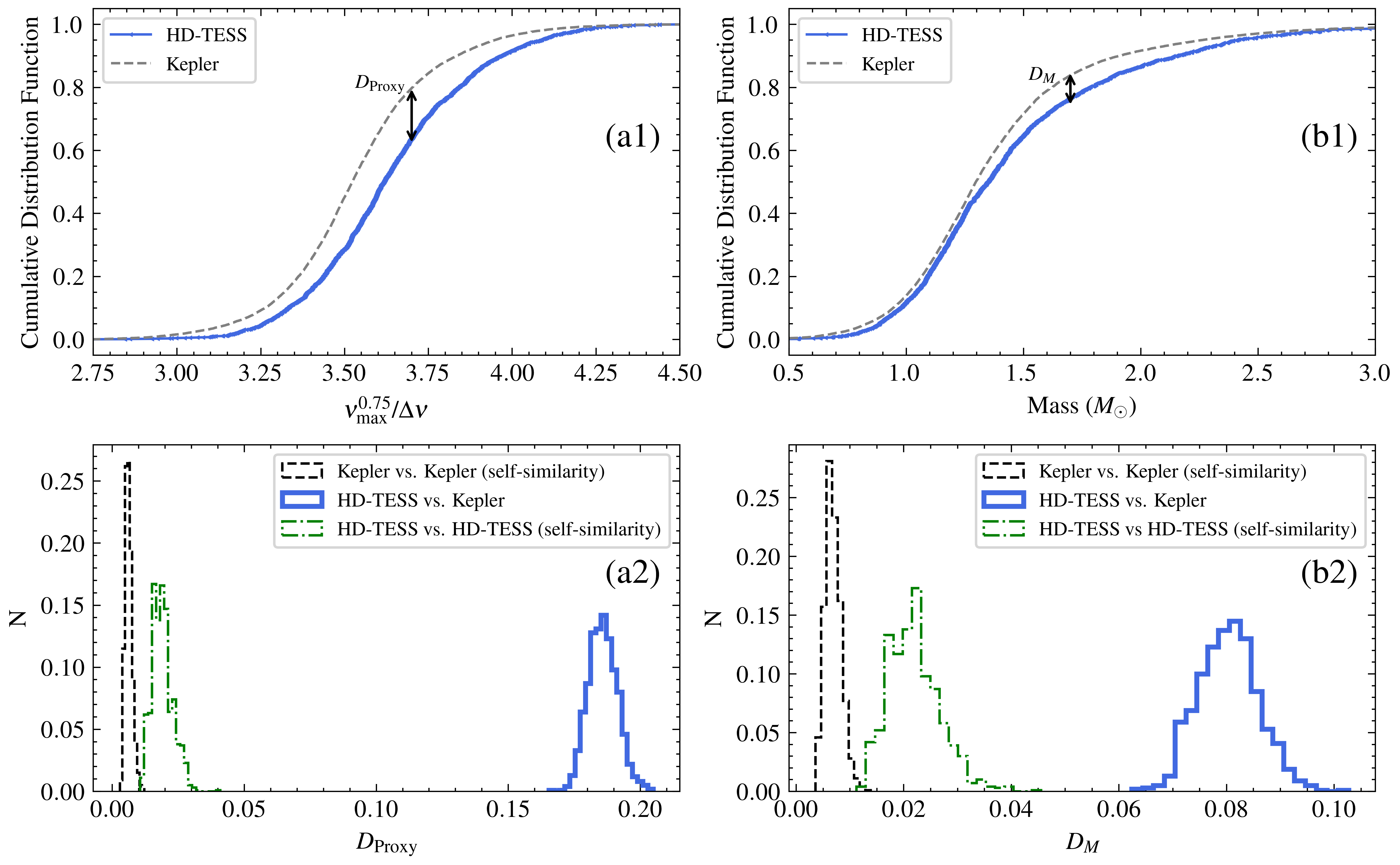}
    \caption{(a1) Empirical cumulative distribution functions (ECDFs) for the HD-TESS and \textit{Kepler} asteroseismic mass proxy values, $\nu_{\mathrm{max}}^{0.75}/\Delta\nu$. The quantity $D$ measures the maximum vertical distance between any two ECDFs. Here, the plotted ECDFs (HD-TESS vs \textit{Kepler}) are constructed using the mean values of $\nu_{\mathrm{max}}$ and $\Delta\nu$. (a2) The distribution of $D_{\mathrm{Proxy}}$ across $10^5$ different noise realizations. The black, dashed histogram is a `self-similarity' $D_{\mathrm{Proxy}}$ distribution for the \textit{Kepler} dataset, created by repeatedly comparing one instance of a \textit{Kepler} ECDF to another having a different noise realization. The dash-dotted green histogram is the self-similarity distribution for the HD-TESS dataset. The blue, solid histogram measures similarity \textit{between} HD-TESS and \textit{Kepler} datasets, and its large offset from the other histograms indicates a strong dissimilarity between the underlying Kepler and TESS seismic mass proxy distributions that is unlikely to be explained by random uncertainties in Kepler or TESS datasets alone. Panels (b1) and (b2) are similar to panels (a1) and (a2), respectively, but for reported values of mass. }
    \label{fig:stats}
\end{figure*}
Given the variety of atmospheric parameter catalogs used, it is difficult to calibrate each to a common temperature or metallicity scale. Although such calibrations can be performed across one large catalog to another (e.g., APOGEE versus \textit{Starhorse}, \citealt{Anders_2022}), spectroscopic catalogs of bright stars such as PASTEL combine measurements across numerous `boutique' surveys, for which systematic differences across each survey are difficult to quantify. We therefore do not make an attempt to calibrate the atmospheric parameters across each catalog to a common scale. Instead, we suggest users to be aware of the origins of the effective temperature and metallicity values used for the mass or radius determination (which we report in Table \ref{table:mass_radii_measurements}), and to be cautious of systematic offsets between temperatures and metallicity scales between different atmospheric parameter catalogs. For consistency, we recommend the use of $\kappa_M$ and $\kappa_R$ from Table \ref{table:seismic_measurements} along with homogenously measured atmospheric parameters.

Table \ref{table:mass_radii_measurements} tabulates mass and radius for each star using Equations \ref{mass_scaling_relation} and \ref{radius_scaling_relation}, with $f_{\Delta\nu}$ computed using the \citet{Sharma_2016} method, which takes into account the temperature, metallicity, and giant evolutionary state. Here, the quoted uncertainties for masses and radii are estimated from 1$\sigma$ intervals from Monte Carlo sampling of the scaling relations assuming no correlations between measurements used as inputs. The resulting median uncertainty across our sample is 8.3\% for mass and 3.2\% for radius, neglecting any potential systematic errors between atmospheric parameter catalogs. 

Figure \ref{fig:kde_mass} compares the distribution of seismic masses of our TESS giants with those from \textit{Kepler} \citep{Yu_2018} where the
TESS sample noticeably contains a greater fraction of stars above $\sim1.5M_{\odot}$. 
Such a result is consistent with the comparisons made using purely using the seismic mass proxy in Figure \ref{fig:nike}, and so these relative mass differences are unlikely to be caused by offsets in temperature and metallicity scales between atmospheric sources. This result suggests that HD-TESS has fractionally more young stars, which is expected of stars in the local neighbourhood for which many have [Fe/H] and [$\alpha$/Fe] abundances at the solar level (e.g., \citealt{Haywood_2013, Hayden_2015}). The lower panel of Figure \ref{fig:kde_mass} shows that the \textit{Kepler} sample comprises more distant stars, many of which belong to older Galactic populations including the thick disk \citep{Silva-Aguirre_2018, Miglio_2021} as well as the Galactic halo \citep{Epstein_2014, Mathur_2016}.

To establish the differences in masses between the \textit{Kepler} and HD-TESS datasets more concretely, we compare their empirical cumulative distribution functions (ECDFs) in Figure \ref{fig:stats}. Panel (a1) shows systematic offsets in the between \textit{Kepler} and TESS the seismic mass proxy ECDFs that manifest as a relative overdensity of stars with $1.2M_{\odot}\apprle M \apprle 2.2 M_{\odot}$ for HD-TESS masses in panel (b1). To include the effects of uncertainties in these comparisons, we apply a Monte-Carlo approach to the $D$-statistic calculated from the two-sided Kolmogorov-Smirnov test \citep{Massey_1951}. The $D$-statistic measures the maximum vertical distance between two ECDFs, with larger values generally indicating greater levels of dissimilarity between the two compared datasets. To quantify the impact of our measurement uncertainties, we first perturb each quantity (seismic mass proxy or stellar mass) from a dataset with its associated uncertainty and construct an instance of its ECDF. A single value of the $D$-statistic is then computed by comparing this ECDF to another constructed using quantities with another realization of noise or uncertainty. This process is repeated $10^5$ times to build up a $D$-statistic distribution as shown in panels (a2) and (b2) of Figure \ref{fig:stats}. The `\textit{Kepler} vs. \textit{Kepler}' and `HD-TESS vs. HD-TESS' distributions in these panels are self-similarity distributions. Specifically, these are created by comparing ECDFs of the same dataset (e.g, \textit{Kepler} with itself) but across different noise realizations to show how much variation in the $D$-statistic is expected from measurement uncertainties alone. Meanwhile, the `HD-TESS vs. \textit{Kepler}' distribution measures similarity between datasets, which shows a large offset compared to the self-similarity distributions. The implication here is that differences in the compared quantities between HD-TESS and \textit{Kepler} datasets are unlikely caused by measurement uncertainties alone, and we thus conclude that the increase in average mass seen for our TESS dataset relative to \textit{Kepler} is indeed significant.

\begin{figure}
    \centering
	\includegraphics[width=1.\columnwidth]{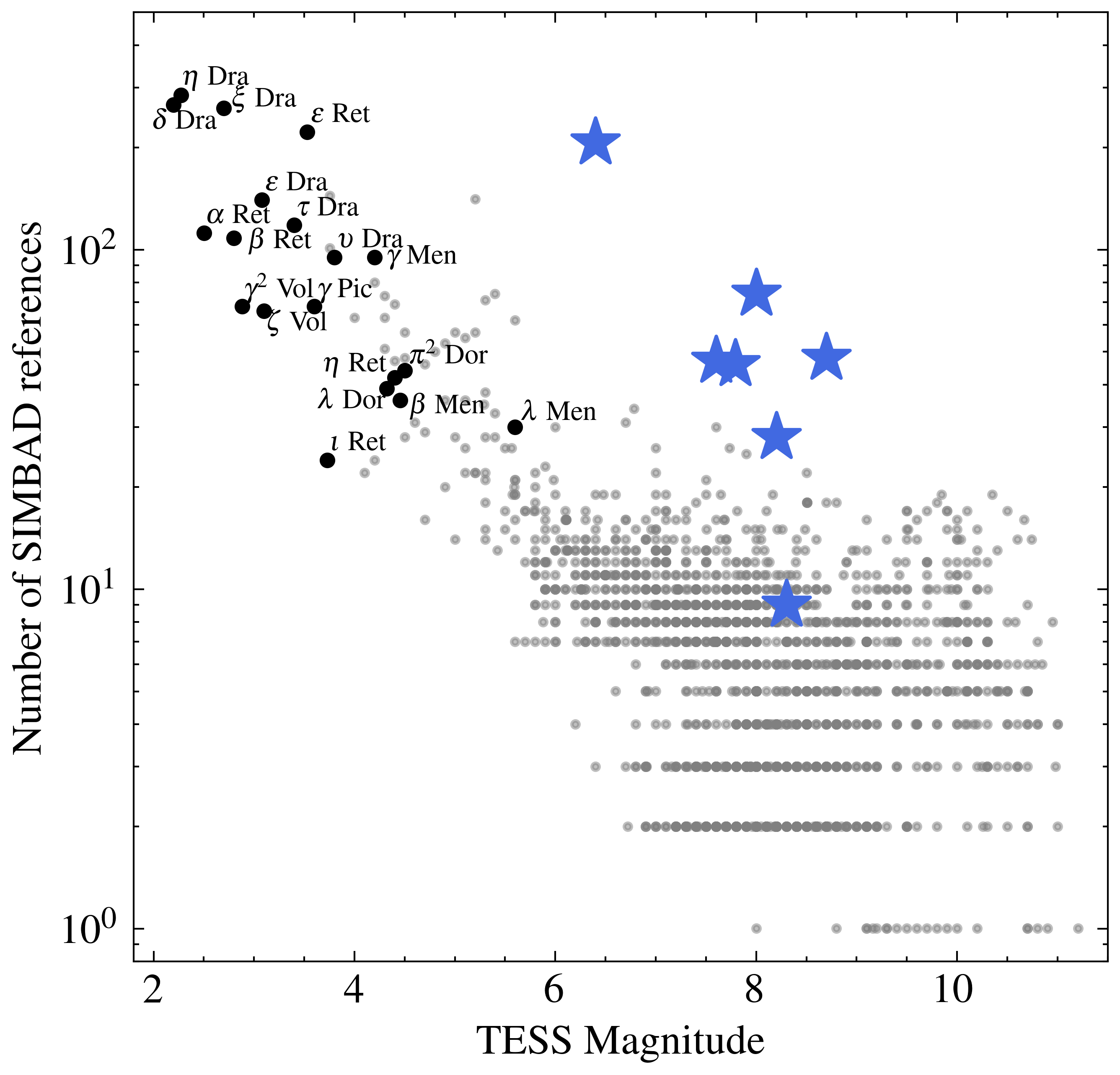}
    \caption{The relative popularity of bright, oscillating red giants in our TESS sample, visualized as the number of references registered on the SIMBAD astronomical database \citep{Wenger_2000} as of January 2022. Bright stars with Bayer designations are annotated in the plot, while blue points indicate interesting giants flagged as `peculiar' by SIMBAD, which here comprises either halo stars or known barium stars \citep{Bidelman_1951}.}
    \label{fig:giraffe}
\end{figure}

\begin{figure*}
    \centering
	\includegraphics[width=2.\columnwidth]{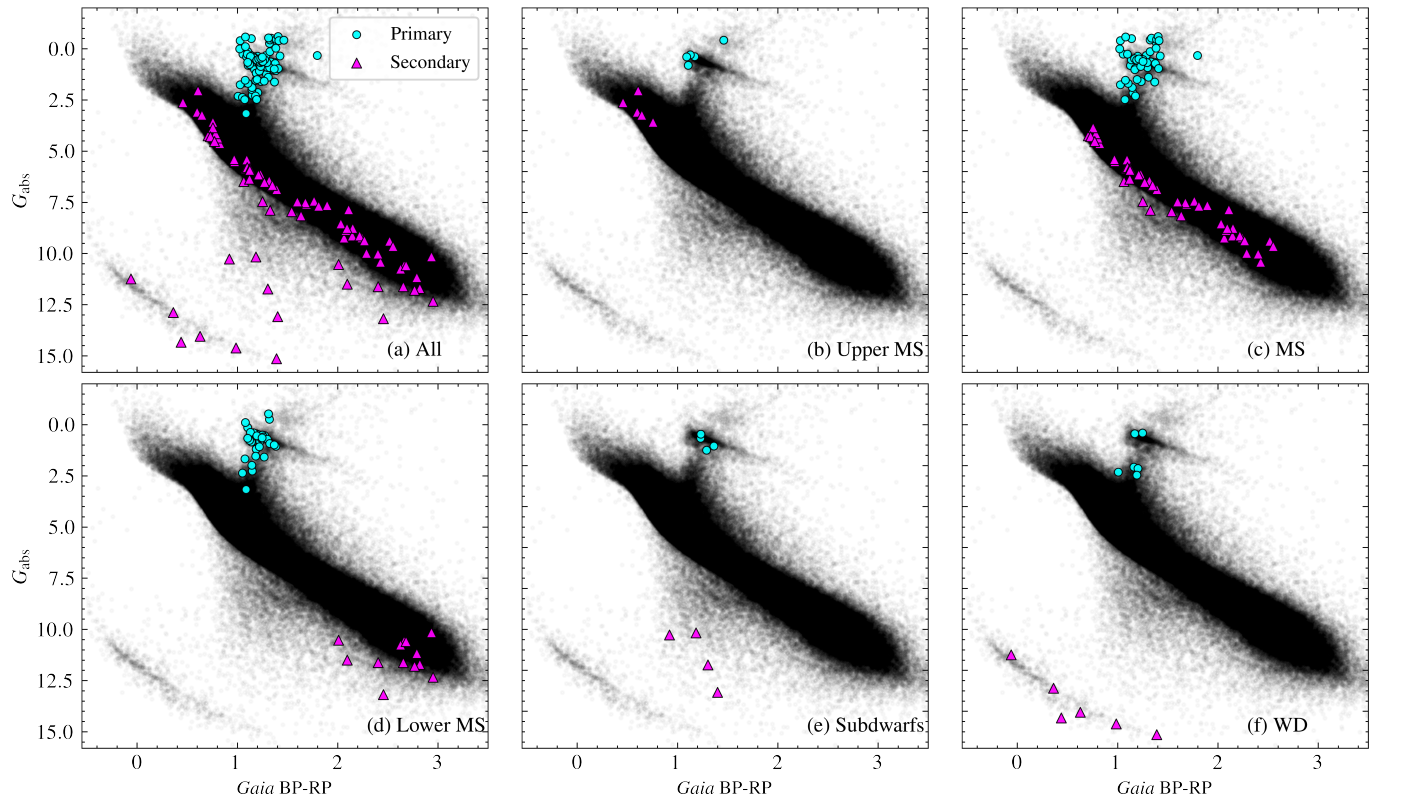}
    \caption{\textit{Gaia} color-magnitude diagrams showing the 99 oscillating giants in our sample known to be astrometric binaries from the \citet{ElBadry_2021} catalog. Only binary systems that have a chance alignment probability less than 0.1 from the catalog are identified. The blue points mark the brighter, primary component (oscillating giant) of each binary system, while the magenta points mark the corresponding fainter, secondary component. The black points are a random subset of stars from \textit{Gaia} EDR3 \citep{GaiaEDR3}. Panel (a) shows all identified binary systems, which are further categorized in panels (b-f) based on the location of their secondary components in the color-magnitude diagram: (b) on the upper main sequence (MS); (c) on the main sequence; (d) on the lower main sequence; (e) in the subdwarf regime; (f) along the white dwarf (WD) sequence. }
    \label{fig:binary}
\end{figure*}

\begin{figure*}
    \centering
	\includegraphics[width=2.05\columnwidth]{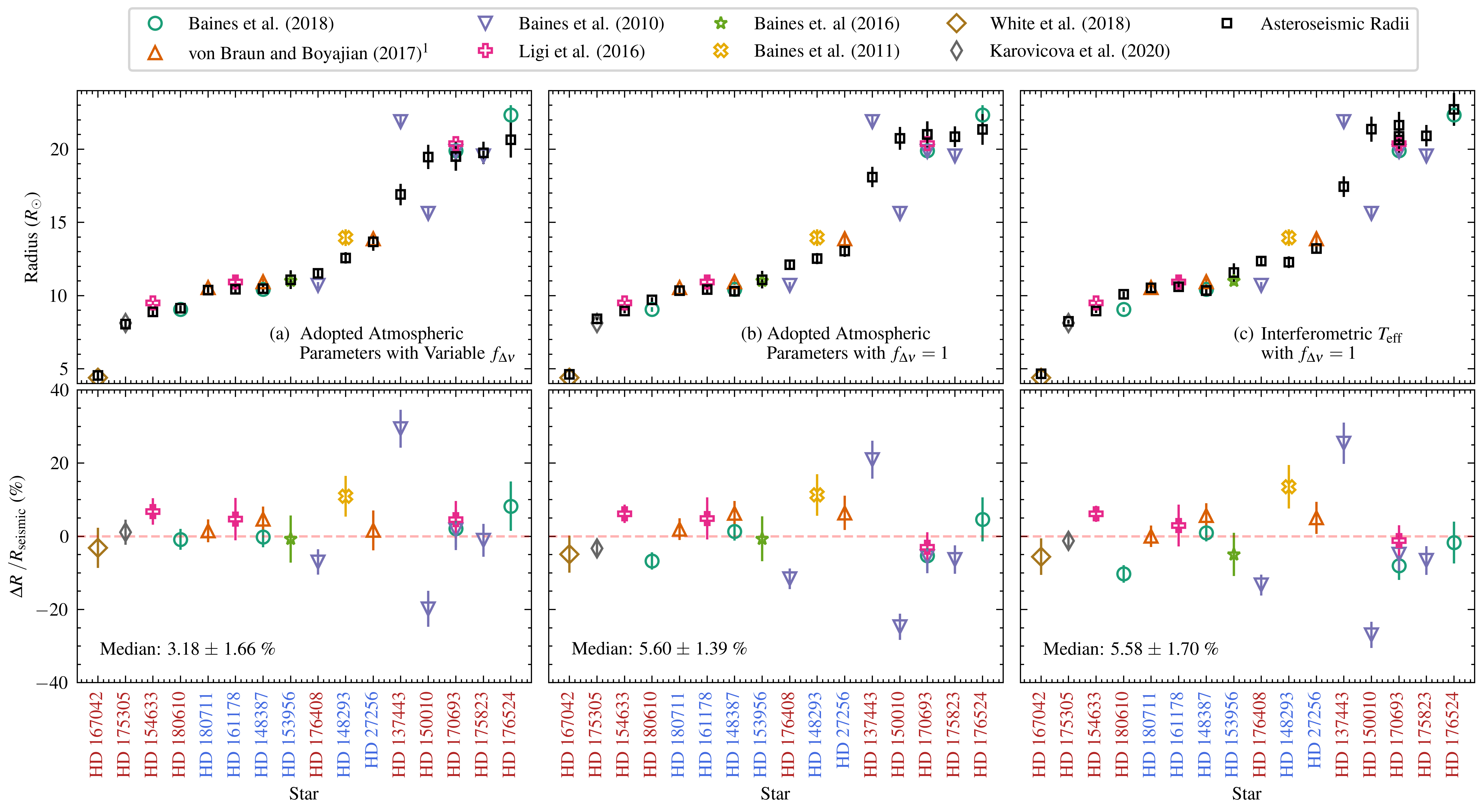}
    \caption{Comparisons of radii from the asteroseismic scaling relations with linear radii from interferometry across literature for 16 oscillating giants from the HD-TESS catalog. The lower panels show fractional difference between radii estimates, where $\Delta R=R_{\mathrm{interferometric}} - R_{\mathrm{seismic}}$. The median $\Delta R/R_{\mathrm{seismic}}$ for each comparison is quoted in each lower panel, with standard errors as the reported uncertainty. The color of each star's label along the x-axis reflects its evolutionary state, with red for RGB and blue for CHeB. (a) $R_{\mathrm{seismic}}$ uses ($T_{\mathrm{eff}}$, [M/H]) adopted in this study and applies the \citet{Sharma_2016} correction to compute the correction factor $f_{\Delta\nu}$ (see Equation \ref{radius_scaling_relation}).
    (b) $R_{\mathrm{seismic}}$ uses $T_{\mathrm{eff}}$ values adopted in this work, with $f_{\Delta\nu}=1$. (c) $R_{\mathrm{seismic}}$ uses the literature (i.e., interferometric) $T_{\mathrm{eff}}$ values reported from each cited work, with $f_{\Delta\nu}=1$.}
    \label{fig:interfero}
\end{figure*}

\section{Scientific Uses for this Catalog}\label{section:science}
Owing to their brightness and intrinsic luminosity, nearby red giants have historically been prime candidates for studies of local stellar populations, such as chemical abundance surveys using high resolution spectroscopy (e.g., \citealt{Hekker_2007, Luck_2007, Luck_2015, Liu_2014, Takeda_2014}) and planet searches using the radial velocity technique (e.g., \citealt[and references therein]{Dollinger_2021}). Beyond local surveys, many of these giants have been individually studied for a wide variety of science cases.
Figure \ref{fig:giraffe} shows that the number of times some of the brightest targets has been included in a publication ranges in the hundreds, while even for fainter targets there still exists a subset of giants ($T_{\mathrm{mag}}\apprge6$) that are of great interest scientifically.
Prior to TESS, however, a vast majority of these giants have not had any precise seismic characterization to complement the studies involving such stars. Although this paper is primarily concerned with assembling a catalog of seismic measurements, we highlight several useful science cases of our catalog unique to bright stars that we will pursue further in future work.

\subsection{Stellar Multiplicity}
The ever increasing precision of astrometry from the \textit{Gaia} space mission \citep{Gaia_2018, GaiaEDR3} has proven extremely valuable for the study of both resolved (e.g., \citealt{El-Badry_2019b, ElBadry_2021, Makarov_2021}) and unresolved stellar companions (e.g., \citealt{Belokurov_2020, Penoyre_2022}) of nearby stars. Of particular interest are binary systems for which one of its components is an oscillating giant because these enable direct constraints on the age of the co-eval companion. In
Figure \ref{fig:binary}a, we show a cross-match of our TESS sample to the catalog of astrometric binaries from \citet{ElBadry_2021}, yielding a total of 99 oscillating giants with a low chance alignment probability ($<0.1$). Figures \ref{fig:binary}b-f show that the stars comprising the secondary component in these binary systems span a variety of evolutionary stages based on the color-magnitude diagram, ranging from parts of the main sequence to white dwarfs.
Because the giants in our sample generally have well-characterized oscillations, this offers the opportunity of calibrating age relations in other parts of the color-magnitude diagram such as gyrochronology \citep{Skumanich_1972, Barnes_2007}, activity indicators along the main sequence (e.g., \citealt{Pace_2013}), and the cooling-age relation for white dwarfs \citep{Winget_1987}. These 99 oscillating giants are tabulated in Appendix \ref{section:appendix_gaia_binaries}.

The orbital solutions of resolved astrometric binaries, measured by methods such as speckle interferometry (e.g., \citealt{Mason_1999, Balega_2002}) or from space astrometry \citep{Makarov_2021, Makarov_2021b}, provide an estimate of the mass ratio of the binary system. In such scenarios, the availability of precise mass constraints for one component (in this case the red giant primary) for such systems translates into a mass determination of both binary components, which is valuable not only for testing stellar evolution theory, but for estimating masses of stars for which model-independent masses are rare such as subdwarfs and white dwarfs. A cross-match of our sample with the Washington Double Star Catalog \citep{Mason_2022} reveals that 91 of our TESS giants are in known double star systems, while a cross-match with the Fourth Catalog of Interferometric Measurements of Binary Stars \citep{Hartkopf_2001} yields 57 stars with previously measured positions. These are tabulated in Appendix \ref{section:appendix_interferometric_binaries}.

\subsection{Benchmarking Asteroseismic Radii with Interferometry}\label{section:interferometry_benchmark}
The direct measurement of stellar radii using interferometry provides a powerful, model-independent approach to validate the accuracy of radii inferred using the asteroseismic scaling relations \citep{Huber_2012, White_2015}. Given known offsets of masses and radii from asteroseismic scaling relations with respect to stellar models (e.g., \citealt{Epstein_2014, Sharma_2016}) and eclipsing binaries (e.g., \citealt{Brogaard_2018}), it is important that comparisons between interferometric radii with seismic radii be performed across a wide range of metallicities, temperatures, and evolutionary stages. The bottleneck, however, has been the availability of precise seismic data for interferometric targets, for which to date only 7 giants with $R\apprge3R_{\odot}$ has had published joint seismic-interferometric coverage \citep{Huber_2012, Johnson_2014, Beck_2015}. The HD-TESS catalog can significantly alleviate this limitation by providing a large quantity of bright targets amenable to both asteroseismology and interferometry. 

Figure \ref{fig:interfero} compares radii for 16 red giants with published interferometric measurements \textemdash{} a factor of two increase from literature. The power spectrum and seismic measurements for each of these 16 giants is presented in Appendix \ref{section:appendix_interf_spectrum}.
The interferometric radii from each literature study (\citealt{Baines_2010, Baines_2011, Baines_2016, Baines_2018}, \citealt{Ligi_2016}, \citealt{vonBraun_2017}\footnote{The following stars reported by \citet{vonBraun_2017} have reported angular diameters that are weighted averages of other literature values: HD 27256 \textemdash{} \citet{Cusano_2012}; HD 148387 \textemdash{} \citet{Nordgren_2001}; HD 180711 \textemdash{} \citet{Nordgren_2001, Mozurkewich_2003}.}, \citealt{White_2018}, \citealt{Karovicova_2020}) are computed by combining reported limb darkened angular diameters with parallaxes from \textit{Gaia} EDR3. In Appendix \ref{section:appendix_hip_parallax}, we perform the same analysis as in this Section using interferometric radii derived but using Hipparcos parallaxes \citep{vanLeeuwen_2007}, where we generally find worse agreement. We thus use \textit{Gaia} EDR3 parallaxes for the rest of the analysis presented here. We correct for potential zero-point offsets in the parallaxes following \citet{Lindegren_2020} and the correction recommended by \citet{Zinn_2021}.

Three comparisons are presented in Figure \ref{fig:interfero}: (a) Using the \citet{Sharma_2016} correction, which provides a $f_{\Delta\nu}$ dependent on the $T_{\mathrm{eff}}$, [Fe/H], and evolutionary state of each star reported in this work; (b) No corrections (i.e., $f_{\Delta\nu}$ = 1) with the adopted $T_{\mathrm{eff}}$ in this work, (c) No corrections and using the $T_{\mathrm{eff}}$ reported from each interferometric study. The latter comparison is made in the interest of consistency as the quoted literature $T_{\mathrm{eff}}$ values are typically derived using angular diameters through an application of the Stefan-Boltzmann law. We observe the following:

\begin{itemize}
    \item The application of a star-dependent correction factor $f_{\Delta\nu}$ generally improves the agreement between seismic and interferometric radii, especially for low-luminosity giants with $R\apprle15R_{\odot}$. In particular, we find the median fractional difference between the corrected seismic radii and interferometic to be approximately at the 3.2 $\pm$ 1.7\% level. In contrast, the use of no correction factor in Figures \ref{fig:interfero}b-c result in greater median fractional differences ($\sim5-6\%$).
    These findings are in accordance with other benchmark studies of asteroseismic radii using \textit{Gaia} parallaxes \citep{Zinn_2019} as well as using dynamical radii from eclipsing binaries \citep{Gaulme_2016, Brogaard_2018, Kallinger_2018}, and further confirms that a correction factor to the asteroseismic scaling relations is indeed warranted. 
    
    \item The similarity of trends between panels (b) and (c) suggests that the improvements seen in (a) are mostly from the adoption of a star-dependent $f_{\Delta\nu}$, rather than by differences in the adopted effective temperatures (spectroscopic versus interferometric). Regardless of the adopted effective temperature, scenarios (b) and (c), which do not adopt a star-dependent $f_{\Delta\nu}$, show larger discrepancies compared to the opposite scenario in (a). This implies that the presence of any form of $f_{\Delta\nu}$ is better than no correction whatsoever.
    
    \item The two most discrepant stars, HD 137443 and HD 150010, are labelled as `RGB' stars with a measured $\nu_{\mathrm{max}}\sim10-15\mu$Hz. It is thus plausible that these are actually helium shell-burning asymptotic giant branch (AGB) stars, which are generally difficult to disentangle from hydrogen shell-burning RGB stars seismically (e.g., \citealt[their Figure 4b]{Stello_2013}, \citealt{Dreau_2021}). The \citet{Sharma_2016} correction does not yet account for evolutionary stages past core-helium burning, which may explain the large observed differences between seismic and interferometric radii. 
    We note, however, that the interferometry for these stars are only based on only a single series of measurements by \citet{Baines_2010}, and thus could be affected by instrument-dependent systematics in angular diameter measurements \citep{Tayar_2022}.

\end{itemize}

\section{Conclusions}
Our main conclusions are as follows:

\begin{enumerate}
    \item We present a catalog of fundamental asteroseismic parameters $\Delta\nu$ and $\nu_{\mathrm{max}}$ and evolutionary states for 1,709 bright, oscillating red giants from the Henry Draper Catalog. These stars and required to have been observed for at least 6 months within one TESS observing cycle throughout Cycles 1-3 and thus are located near the TESS continuous viewing zones.
    \item We provide seismic mass ($\kappa_M$) and radius ($\kappa_R$) coefficients for users that wish to use their own effective temperatures and metallicities when applying the asteroseismic scaling relations to estimate masses and radii. With the application of such scaling relations, we determine masses with a typical precision of 8.3\% and radii with a typical precision of 3.2\% for the stars in our sample using a compilation of atmospheric parameters from literature. Compared to the population of \textit{Kepler} red giants, the HD-TESS sample is systematically more massive, with a larger fraction of stars having $M\apprge1.5M_{\odot}$.
    \item 99 stars in our sample are known astrometric binaries from \textit{Gaia} EDR3, while 40 are well-known visual binaries with archival measurements from the Washington Double Star catalog. These binaries include those with subdwarfs and white dwarfs as the secondary.
    \item We benchmark radii from asteroseismic scaling relations against radii measured using interferometry for 16 giants in our sample. Our results show that the median fractional difference between the two radii are approximately 3\% for giants with $R<15R_{\odot}$ when calculating interferometric radii using \textit{Gaia} parallaxes. Reaching this level of agreement between the two approaches, however, requires a correction factor to the scaling relations, in agreement with other independent benchmarking studies.

\end{enumerate}

The preliminary work here highlights the value of TESS observations in calibrating methods of inferring fundamental stellar parameters of well-studied bright stars. The large number of precise seismic measurements from this work, which when combined with accurate \textit{Gaia} parallaxes and ground-based spectroscopy,
will allow more frequent and consistent benchmarks of asteroseismic inference methods against interferometry.
Neither the highlighted science cases nor the seismic measurements provided in this work are exhaustive in nature. As TESS pursues further observing cycles, not only will the asteroseismic precision of frequently visited regions of the sky increase, but more bright stars near us \textemdash{} particularly those closer to the ecliptic plane \textemdash{} will also eventually be accessible with precision asteroseismology. In that sense, the work presented here acts as a precursor for a more comprehensive asteroseismic catalog of bright stars in the near future, where we will aim for a much more extensive sky coverage.


\vspace{5mm}
\facilities{TESS}


\appendix

\section{Frequency at Maximum Power for 2,611 Oscillating Giants}\label{section:appendix_numax}
Table \ref{table:numax} lists the $\nu_{\mathrm{max}}$ measurements for 2,611 oscillating giants from the HD catalog that show oscillations in this work. The table entries comprise the 1,709 stars forming the main HD-TESS catalog (Table \ref{table:seismic_measurements}) as well as 903 additional red giants whose $\Delta\nu$ cannot be measured in this study.

\begin{deluxetable}{ccc}[t]
\tablenum{A1}
\tablecaption{The measured frequency of maximum power, $\nu_{\mathrm{max}}$ of 2,611 giants with detected oscillations in this work. Each star's identifier in the Henry Draper (HD) catalog and the TESS Input Catalog (TIC) is provided. The full version of this table is available in a machine-readable format in the online journal, with a portion shown here for guidance regarding its form and content. \label{table:numax}}
\tablewidth{0pt} 
\tablehead{ 
\colhead{HD number} & \colhead{TIC identifier} & \colhead{$\nu_{\mathrm{max}}$}
\\
\colhead{} & \colhead{} & \colhead{$\mu$Hz}
}
\startdata
22349 & 31852879 & $34.04\pm0.50$\\
23128 & 238180822 & $25.24\pm0.39$\\
23253 & 31959761 & $55.80\pm1.25$\\
23266 & 311925612 & $28.29\pm0.75$\\
$\cdots$&$\cdots$&$\cdots$\\
311763 & 272472375 & $18.67\pm1.50$\\
370635 & 370115759 & $22.95\pm0.35$\\
\enddata
\end{deluxetable}

\section{Spatially Resolved Astrometric Binaries from \textit{Gaia} EDR3}\label{section:appendix_gaia_binaries}
Table \ref{table:gaia_astrometric_binaries} lists the 99 oscillating giants in our sample that have been identified as being part of an astrometric binary system in the \citet{ElBadry_2021} catalog. From the cross-match, we select only binaries with a computed chance alignment probability less than 0.1.

\begin{deluxetable}{ccc}[t]
\tablenum{B1}
\tablecaption{List of 99 binaries from the \citet{ElBadry_2021} catalog for which the primary is an oscillating giant in our sample. The quoted \textit{Gaia} source IDs are those from Early Data Release 3 \citep{GaiaEDR3}. The full version of this table is available in a machine-readable format in the online journal, with a portion shown here for guidance regarding its form and content. \label{table:gaia_astrometric_binaries}}
\tablewidth{0pt} 
\tablehead{ 
\colhead{HD number} & \colhead{Primary Source ID} & \colhead{Secondary Source ID}
\\
}
\startdata
24272 & 4667870070470785792 & 4667870070470786816\\
27068 & 4676770239141111424 & 4676770239141111040\\
27670 & 4676012709989349376 & 4676012778708825344\\
28633 & 4656569496121672832 & 4656569461761934592\\
$\cdots$&$\cdots$&$\cdots$\\
311373 & 5215069141868112256 & 5215069141868555648\\
311734 & 5263024685110990720 & 5263024685110990592\\
\enddata
\end{deluxetable}

\section{Double Stars and Interferometric Binaries}\label{section:appendix_interferometric_binaries}
Table \ref{table:wds} lists a cross-match of our oscillating giant sample to the Washington Double Star catalog \citep{Mason_2022}. Table \ref{table:interf_binary} presents a cross-match to the Fourth Catalog of Interferometric Measurements of Binary Stars \citep{Hartkopf_2001}. The full version of both tables are available in a machine-readable format in the online journal, with a portion shown here for guidance regarding their form and content.

\begin{deluxetable}{cc}[t]
\tablenum{C1}
\tablecaption{List of 91 oscillating giants that are known double stars in the Washington Double Star (WDS) Catalog.  \label{table:wds}}
\tablewidth{0pt} 
\tablehead{ 
\colhead{HD number} & \colhead{WDS identifier} 
\\
}
\startdata
23817 & WDS J03442-6448A \\
24461 & WDS J03480-7048A \\
25106 & WDS J03549-6627AB \\
27256 & WDS J04144-6228A \\
$\cdots$&$\cdots$\\
239229 & WDS J19490+5916A\\
269894 & WDS J05385-7045AB\\
\enddata
\end{deluxetable}

\begin{deluxetable}{cc}[t]
\tablenum{C2}
\tablecaption{List of 57 oscillating giants with archival position measurements from the Fourth Catalog of Interferometric Measurements of Binary Stars.  \label{table:interf_binary}}
\tablewidth{0pt} 
\tablehead{ 
\colhead{HD number} & \colhead{TIC ID}
\\
}
\startdata
132464 & 219018004\\ 
132698 & 288185404\\
137292 & 219731897\\
141264 & 272746493\\
$\cdots$\\
238629 & 199713797\\
270011 & 389325554\\
\enddata
\end{deluxetable}

\section{Power Spectra of Oscillating Giants with Interferometry}\label{section:appendix_interf_spectrum}
\begin{figure}
    \centering
	\includegraphics[width=1.0\columnwidth]{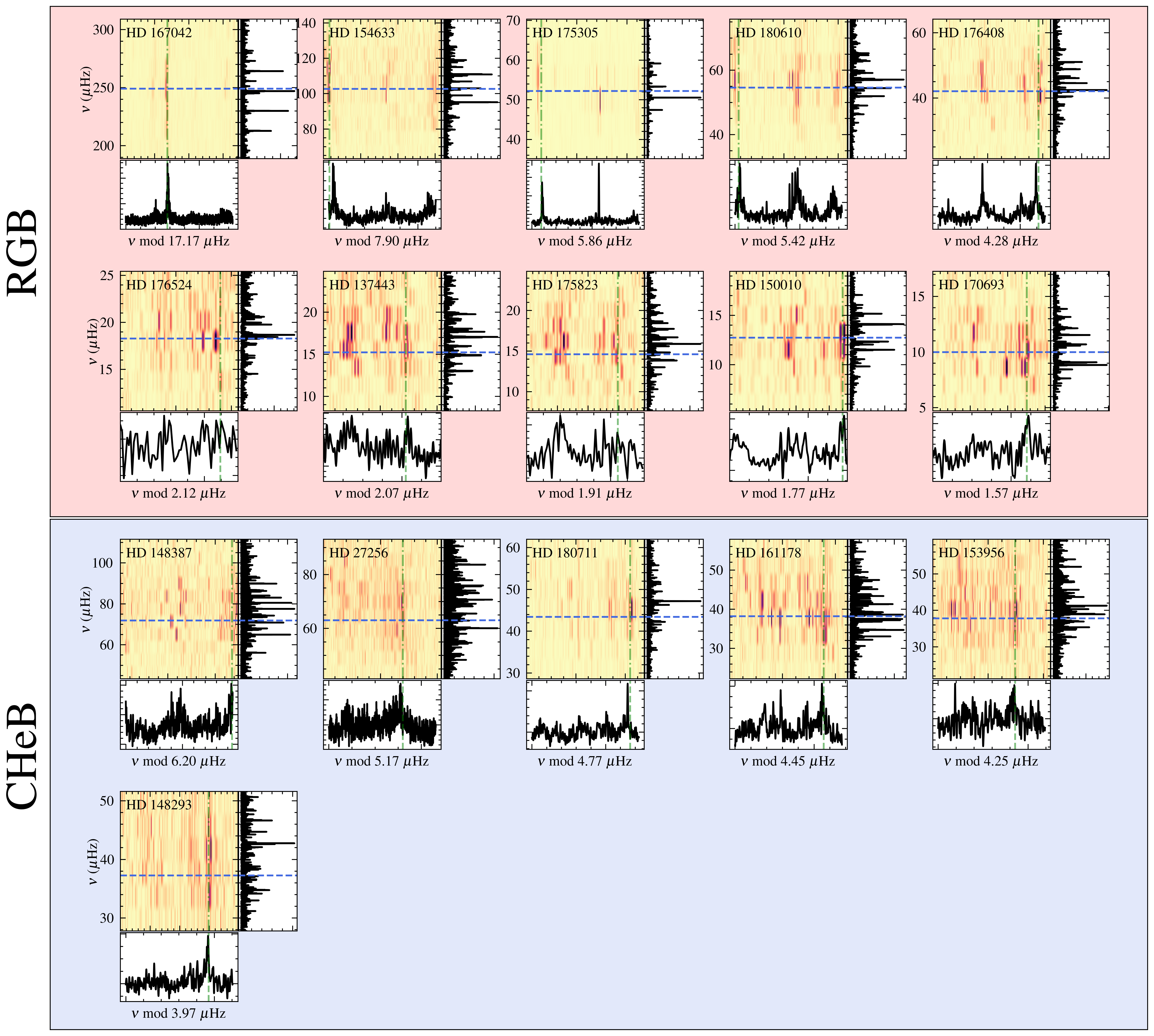}
    \caption{Seismic visual diagnostic plots for the 16 oscillating giants benchmarked against interferometric measurements in Section \ref{section:interferometry_benchmark}, categorized by their evolutionary state (RGB = hydrogen shell-burning, CHeB = core helium-burning). The stars are ordered in decreasing $\Delta\nu$ from the top left to the bottom right in each category. The blue horizontal line in each panel corresponds to $\nu_{\mathrm{max}}$, while the green vertical line corresponds to the location of the radial mode-ridge in the \'echelle diagrams.}
    \label{fig:interfero_spectrum_appendix}
\end{figure}

Figure \ref{fig:interfero_spectrum_appendix} shows the \'echelle diagram and power spectra of the 16 oscillating red giants with interferometric radii discussed in Section \ref{section:interferometry_benchmark}.

\section{Seismic Radii versus Interferometric Radii Computed using Hipparcos Parallaxes}\label{section:appendix_hip_parallax}
\begin{figure}
    \centering
	\includegraphics[width=1.0\columnwidth]{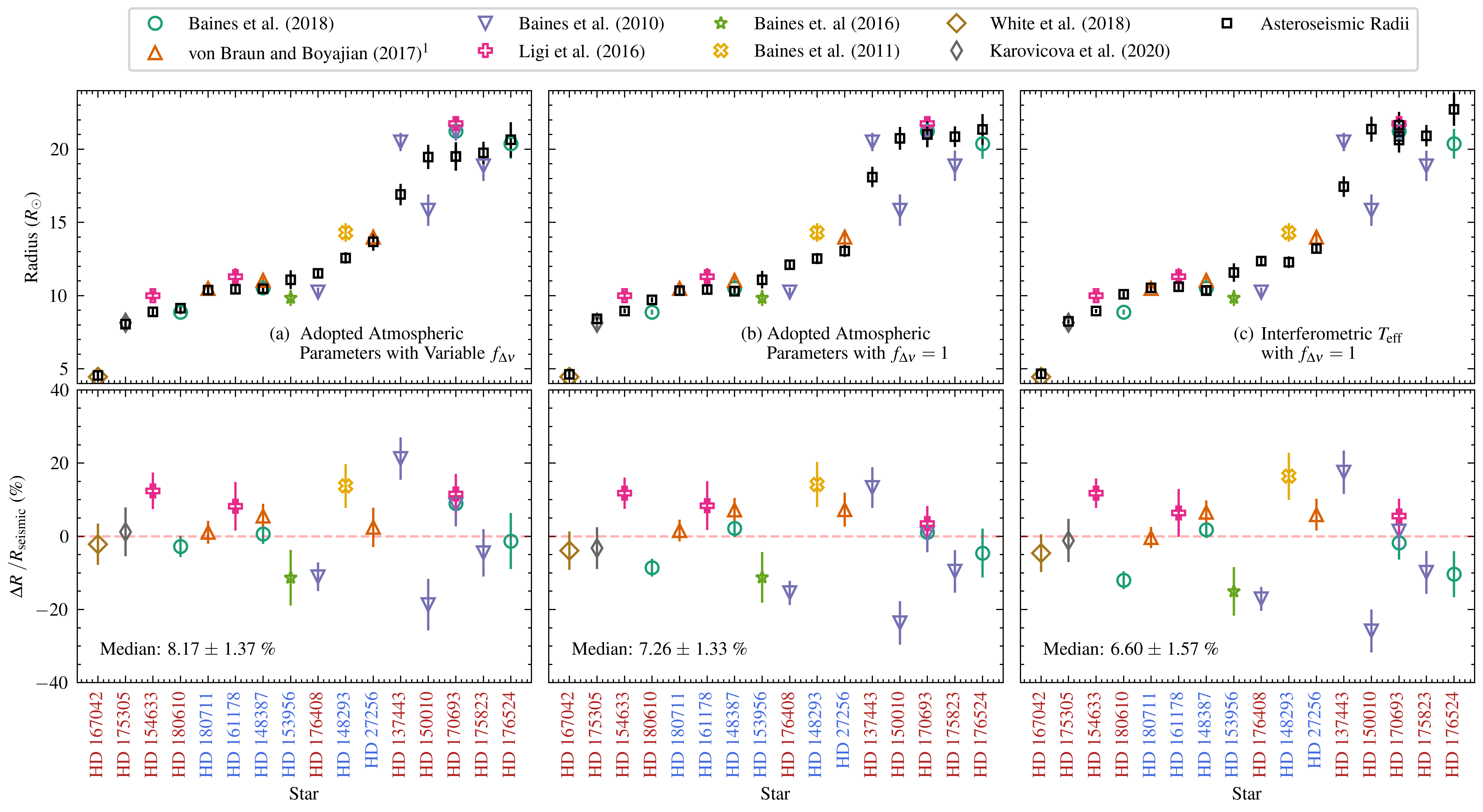}
    \caption{Comparisons of radii from the asteroseismic scaling relations with linear radii from interferometry across literature for 16 oscillating giants from our TESS giant sample. This Figure is identical to Figure \ref{fig:interfero}, except that the inteferometric radii are derived using Hipparcos parallaxes as opposed to \textit{Gaia} EDR3 parallaxes.}
    \label{fig:interfero_appendix}
\end{figure}

Figure \ref{fig:interfero_appendix} repeats the analysis in Section \ref{section:interferometry_benchmark} but uses interferometric radii derived from Hipparcos \citep{vanLeeuwen_2007} parallaxes instead of \textit{Gaia} EDR3. The results shown here do not change the conclusions from Section \ref{section:interferometry_benchmark}; however the overall worse agreement suggests that \textit{Gaia} parallaxes are more accurate as well as more precise for the giants analyzed in this study.

\bibliography{sample631}{}
\bibliographystyle{aasjournal}



\end{document}